\documentclass[12pt]{iopart}

\usepackage{iopams}
\usepackage{graphicx} 
\usepackage{epstopdf}
\usepackage{cite}
\usepackage{xcolor}
\newcommand{\dbar} {\ensuremath{\,\mathchar'26\mkern-12mu d}}

\begin{document}

\title[Critical properties in harmonic confinement]{Critical properties of weakly interacting Bose gases as modified by a harmonic confinement}

\author{I Reyes-Ayala, F J Poveda-Cuevas, J A Seman, and V Romero-Roch\'{i}n}

\address{Instituto de F\'{i}sica, Universidad Nacional Aut\'{o}noma de M\'{e}xico, Apartado Postal 20-364, 01000 Ciudad de M\'{e}xico, Mexico}

\ead{romero@fisica.unam.mx}

\vspace{10pt}
\begin{indented}
\item[]\today
\end{indented}

\begin{abstract}
The critical properties of the phase transition from a normal gas to a BEC (superfluid) of a harmonically confined Bose gas are addressed with the knowledge of an equation of state of the underlying homogeneous Bose fluid. It is shown that while the presence of the confinement trap arrests the usual divergences of the isothermal compressibility and heat capacities, the critical behavior manifests itself now in the divergence of derivatives of the mentioned susceptibilities. This result is illustrated with a mean-field like model of an equation of state for the homogeneous particle density as a function of the chemical potential and temperature of the gas. The model assumes the form of an ideal Bose gas in the normal fluid while in the superfluid state a function is proposed such that, both, asymptotically reaches the Thomas-Fermi solution of a weakly interacting Bose gas at large densities and low temperatures and, at the transition, matches the critical properties of the ideal Bose gas. With this model we obtain the {\it global} thermodynamics of the harmonically confined gas, from which we analyze its critical properties. We discuss how these properties can be experimentally tested.
\end{abstract}


\vspace{2pc}

\submitto{\JSTAT}

\maketitle

\section{Introduction}

The transition that a {\it homogeneous} interacting Bose gas suffers, from a normal gas to a  Bose-Einstein superfluid condensate (BEC), has been determined to be a continuous critical second order phase transition. Because the order parameter is the complex macroscopic wavefunction of the condensate, it is expected that the transition belongs to the 3D XY universality class, just as the superfluid phase transitions in $^{3}$He and $^{4}$He  \cite{Fisher-rpp30, Weichman-prb33}. An important feature, however, in the current experimental realizations of this transition in ultracold atomic vapors, is the fact that the phenomenon occurs inside confinement traps. This gives rise to fluids with {\it inhomogeneous} density distributions which, in turn, modify their thermodynamic properties with respect to the homogeneous ones \cite{Bagnato-pra35, Dalfovo-rmp99,Pethick-book2002,Romero-Rochin-prl94, Romero-Rochin-bjp35, Sandoval-Figueroa-pre78}. That the trap alters or modifies the thermodynamics of the transition can already been seen in an {\it ideal} Bose gas, where the condensate fraction scales with temperature differently whether the gas is uniform or if it is confined by a harmonic trap \cite{Pethick-book2002}. The main issue we address here is that, in general, for interacting gases, the critical properties of a confined inhomogeneous Bose gas are modified with respect to its homogeneous counterpart. Although our explicit analysis is performed by means of a heuristic mean-field like equation of state, here introduced, the main result is that while the isothermal compressibility and heat capacities show divergences at the critical transitions in the homogeneous case, the analogous thermodynamic susceptibilities in the confined inhomogeneous version do not diverge themselves. Rather, their non-analytic divergences appear now in their temperature derivatives. We argue that this is a modification of the underlying universality class of the homogeneous fluid, caused by the presence of the external potential of the trap.\\ 

The above results follow from the use of the appropriate thermodynamic treatment of a fluid confined by an external inhomogeneous potential. In such a case, the thermodynamics of the inhomogeneous systems can be cast in the framework of the so-called {\it global} thermodynamics, in which the usual volume $V$ and constant hydrostatic pressure $p$ of homogeneous systems are replaced by the own mechanical variables of the trapped fluid, called the global volume and pressure ${\cal V}$ and ${\cal P}$ \cite{Romero-Rochin-prl94, Romero-Rochin-bjp35, Sandoval-Figueroa-pre78}. While the thermodynamics of the uniform and non-uniform versions of the same gas can be directly obtained by the calculation of their respective free energies, one can also find their connection through the local density approximation (LDA) \cite{Dalfovo-rmp99,Pethick-book2002}, which essentially tells us that an inhomogeneous fluid can be decomposed into locally uniform fluids, each with its own effective local chemical potential. This route should allow us to analyze the modification of the critical properties of a given fluid in its uniform state, when the same fluid is then confined by an external trap - a harmonic one in our case. However, a necessary condition to perform this analysis, is the knowledge of the equation of state of the homogeneous fluid, explicitly showing the continuous second-order phase transition. The further use of LDA and global thermodynamics should yield the corresponding critical properties of the confined fluid.\\

The main difficulty in the program described above is the lack of the actual equation of state of an interacting Bose fluid. A weakly interacting fluid can be correctly described near $T = 0$ with Bogoliubov theory \cite{Bogoliubov-jpussr47}, or extensions such as those in Refs. \cite{Bijlsma-pra55} or \cite{Yukalov-pra74}, and it can also be addressed near the critical temperature with a field theoretic renormalization group approach \cite{Amit-book1984, Ma-book2000}. However, there are no explicit equations matching those two limits. Additionally, if such an equation were known, it should describe the critical transition within the universality class of the 3D XY-model. The knowledge of such an equation remains as one of the most challenging theoretical tasks of this field of research. Lacking, however, such an equation of state, does not prevent us from communicating our main results. Certainly, one must resort to an approximate equation of state that, while being imprecise at the transition, must bear all the requirements imposed by thermodynamics in order to correctly describe the physics involved. Our proposed equation of state is by no means arbitrary, but motivated and suggested by the typical fittings performed on the density profiles experimentally obtained in ultracold vapors. But before entering into the details of the proposed equation of state, it is of relevance to advance that the critical properties of such an equation belong to the universality class of the Spherical model. Let us briefly recall the main features of the 3D XY and the Spherical models. The former refers to $N$ two-component spins, $\vec s^{(i)} = s_x^{(i)} {\bf x} + s_y^{(i)} {\bf y}$, such that $\vec s_i \cdot \vec s_i = 1$, while the latter describes $N$ spins $s_j$ that can take any value but that in every configuration it must be obeyed that $\sum_j s_j^2 = N$. Both show a second order phase transition, but the Spherical model can be solved exactly in the limit $N \to \infty$ \cite{Berlin-pra86}. Long ago, it was shown that the ideal Bose-Einstein condensation belongs to the universality class of the Spherical model \cite{Gunton-pr166, Hall-jsp75}. For reference, the specific heat and correlation length critical exponents, $C - C_{T_c} \sim |T - T_c|^{-\alpha}$  and $\xi \sim |T - T_c|^{-\nu}$, have the values $\alpha = -1$  and $\nu = 1/2$. The other exponents can be found with the usual exponents equalities \cite{Fisher-rpp30, Amit-book1984, Ma-book2000}. On the other hand, the superfluid phase transition in $^{4}$He has been shown to experimentally agree \cite{Lipa-prl51} with the critical exponents of the 3D XY model, calculated with renormalization group methods \cite{Burovski-prb74}, and given by $\alpha = - 0.0127...$ and $\nu = 0.6717...$ . Because of their physical similarities to $^{4}$He, it is expected that BEC in vapors of atomic Bose gases, such as $^{87}$Rb, also belong to this universality class. Yet, the important result for us here is that, although the 3D XY and the Spherical model differ in their precise numerical predictions and in their relationship to actual experimental systems, their physical essence is the same, both describe a critical phase transition from a normal fluid to a macroscopic condensate quantum phase.\\

Returning to the characteristics of the here proposed equation of state for the homogeneous fluid, we insist in its motivation by the experimental fittings to density profiles of ultracold gases \cite{Ketterle-proceeding1999, Szczepkowsk-revsciinstr80}. The experiments clearly show that, when BE condensation is reached, the thermal part of the cloud is fitted quite well by a classical ideal gas, while the condensate peak allows for a Thomas-Fermi (TF) adjustment \cite{Dalfovo-rmp99,Pethick-book2002}. Certainly, the matching transition region cannot be well described by these fittings. Hence, following these observations, our proposed model is constructed such that, on the one hand, asymptotically fits, both, an ideal gas in the deep thermal cloud and a TF form near the center of the trap, and on the other hand, it shows proper critical behaviour at the transition. In particular, we build the model such that the isothermal compressibility $\kappa_T$ in the uniform case shows the characteristic divergence $\kappa_T \sim |T - T_c |^{-\gamma}$, with $\gamma$ a critical exponent. As it is known from the theory of critical phenomena \cite{Fisher-rpp30, Amit-book1984, Ma-book2000} this reflects the underlying long-range density correlations at criticality. Our model is then formulated as a {\it quantum} ideal Bose gas for high temperatures, $T \ge T_c$, which belongs to the Spherical model universality class as already mentioned, then, for low temperatures $T \le T_c$, we propose a minimalistic model that shows the same divergence of the isothermal compressibility but that approaches Thomas-Fermi far from it. The model also ensures that temperature, chemical potential and hydrostatic pressure are continuous at the transition. This model thus shows typical critical behavior of the uniform fluid and allows, in a simple and analytical way, to follow the program of global thermodynamics to study the critical properties of the confined inhomogeneous fluid. \\

Although our model can be used to obtain reasonable fits of actual experimental data \cite{Nascimbene-njp12}, as we show in Section 2, it is still a mean-field model. Hence,  part of our goals is to motivate experiments dedicated to obtain high-resolution in-situ density profiles which, by being directly analyzed with LDA and global thermodynamics, could be used ``in reverse'' to retrieve the critical properties of the real fluid without a precise knowledge of the equation of state of a truly 3D XY model. We point out here that experimental studies on critical properties of $^{87}$Rb condensates using global thermodynamics have already been performed \cite{Henn-nucphysa790,Romero-Rochin-pra85, Shiozaki-pra90, Poveda-Cuevas-pra92, Castilho-njphys18}, however, those experiments have not addressed the specific issues we discuss here. Moreover, in general, an important technical difficulty is the obtention and processing of in-situ density profiles. This technique has indeed already been used in Bose $^7$Li atomic gases \cite{Nascimbene-njp12,Navon-prl107} and in studies of thermodynamics of Fermi superfluids mainly in the unitary regime \cite{Nascimbene-nature463, Ku-Science335}, thus, the global thermodynamic analysis that we propose here can, in principle, be performed. There are also other recent experimental techniques  \cite{Duarte-prl114, Kaminski-ejpd66, Gajdacz-revsciins84, Ramanthan-revsciins83, Wigley-oplett41, Wilson-pra91} that could be used to generate images with the qualities required for the study that we propose here. These techniques have not been used to conduct thermodynamic experiments yet, and therefore we hope our work serves as a motivation for groups with these technical capabilities to explore in detail the critical properties of the Bose gas. We also mention here that recently it has been possible to produce quantum gases in optical traps that resemble ``box" potentials of rigid walls \cite{Schmidutz-prl112, Navon-science347,Chomaz-natcomm6, Mukherjee-prl118}. These experiments certainly simplify the thermodynamic analysis, however, considering the enormous amount has been done, and is still done, in harmonic potentials, we consider relevant to analyze the own thermodynamics of these systems.\\

We organize the article as follows. Section II describes our model of the equation of state of a uniform Bose gas that explicitly shows a critical transition. Section III makes a very brief summary of pertinent global thermodynamics results. Section IV presents analytical solutions of the above mentioned global thermodynamic susceptibilities and, in particular, we discuss their critical behavior. We close with some remarks emphasizing the potential experiments that could be performed and discussing open questions regarding the critical behavior of confined fluids.

\section{An equation of state for a homogeneous Bose gas}

As described in the Introduction, we propose here a model for the equation of state $n = n(\mu, T)$ for a homogenous Bose gas, such that it yields a second order phase transition from a normal gas to a superfluid one in an ultracold gas. The critical properties of the equation of state belong to the universality class of the Spherical model. As such, within the normal gas part it behaves as an ideal Bose gas, while for the superfluid region we propose an equation of state that matches both the density and the fluctuations of the normal part at criticality, but then approaches the Thomas-Fermi behaviour in the deep superfluid side.  \\

To be precise, following the {\it ideal} BEC transition, we assume that for fixed temperature, the phase transition occurs at a critical density value given by the condition that the chemical potential becomes zero, being negative in the normal side. The zero value at the transition, however, may not be the case in a true experiment. That is,  since Bogoliubov theory \cite{Bogoliubov-jpussr47} and TF approximation show a positive chemical potential near zero temperature, the chemical potential must change sign and, thus, the transition could occur at either positive or negative chemical potential and not necessarily at zero.  Recent theoretical evidence has suggested that the transition does occur at negative chemical potential values \cite{Mendoza-Lopez-rmf16}, yet there is no experimental evidence to support any definite value. In any case, our model is as follows. First, the density in the normal region, as a function of chemical potential $\mu$ and temperature $T$ is of the form of the Bose ideal gas,
\begin{equation}
n(\mu, T) = \frac{1}{\lambda_T^3} g_{3/2}(\mu/kT) \>\>\>  {\rm if} \>\>\> \mu \le 0 \label{muneg}
\end{equation}
and, for the superfluid states, we introduce the following heuristic equation of state,
\begin{equation}
n(\mu,T) = \frac{1}{g} \sqrt{(\mu+b g)^{2}- \left(g b \right)^{2}}+n_{c}(T),
  \>\>\>  {\rm if}   \>\>\>  \mu \ge 0 . \label{mupos}
\end{equation}
In Eq. (\ref{muneg}), $\lambda_T = h/(2 \pi mkT)^{1/2}$ is the thermal de Broglie wavelength with $h$ Planck constant, $k$ Boltzmann constant, $m$ the atomic mass, and $g_{3/2}(\mu/kT)$ the Bose function,
\begin{equation}
g_n({\alpha}) = \frac{1}{\Gamma (n)}\int_0^\infty \frac{x^{n-1}}{e^{x-\alpha} - 1} dx .\label{g}
\end{equation}
As we know \cite{Bogoliubov-jpussr47}, the superfluid nature of a Bose gas cannot be described without the presence of atomic interactions. Thus, in the model given by Eq. (\ref{mupos}), for $\mu \ge 0$, $g = 4 \pi \hbar^2 a_s/m$ is the atomic contact interaction, with $a_s > 0$ the $s$-wave scattering length. The quantity $n_c(T)$ is the critical density at fixed temperature $T$ and $\mu = 0$, given by,
\begin{equation}
 n_c(T) = \frac{1}{\lambda_T^3} \zeta(3/2) ,\label{nc}
\end{equation}
with $\zeta(x)$ the Riemann zeta function. This choice ensures the continuity of the density $n$ at the critical value $\mu = 0$, Eqs. (\ref{muneg}) and (\ref{mupos}). The parameter $b$ is a function of $T$, $b = b(T)$, with units of particle density, to be fixed below. A requirement, however, is that $b(T) \to 0$ as $T \to 0$, such that the density takes on the Thomas-Fermi (TF) value for $\mu$ fixed and $T \to 0$, or for $T$ fixed and $\mu \to \infty$,  
\begin{equation}
n(\mu,T) \to \frac{\mu}{g}  \>\>\>{\rm for}\>\>\> \mu \to + \infty  \>\>\>{\rm or}\>\>\> T \to 0. \label{TF}
\end{equation}
That is, the asymptotic form of the present model is in agreement with the fact that the TF model is strictly valid at $T = 0$ only. It may be relevant to indicate that if the interaction $g$ vanishes, the chemical potential must remain zero for $n \ge n(T_c)$ and the density would simply be $n_c(T)$ for $T < T_c$. \\

The main motivation for the present equation of state is that it shows a divergent isothermal compressibility $\kappa_T$ at the transition, as expected from general consideration of the theory of critical phenomena. This is in contrast with previous models, as we review further below. For our purposes we recall the following expression for the isothermal compressibility,
\begin{equation}
\kappa_T = \frac{1}{n^2} \left(\frac{\partial n}{\partial \mu}\right)_T .\label{kappaT}
\end{equation}
The model given by Eqs. (\ref{muneg}) and (\ref{mupos}) shows a divergent first derivative of $n$ with respect to $\mu$ at criticality. This can be seen by expanding both equations (\ref{muneg}) and (\ref{mupos}) near $\mu = 0$. One finds,
\begin{equation}
\left.\frac{\partial n}{\partial \mu}\right|_{\mu \rightarrow 0^{-}} \simeq \frac{\sqrt{\pi}}{\lambda^3 k T} \left| \frac{k T}{\mu} \right|^{1/2},\label{04}
\end{equation}
and
\begin{equation}
\left.\frac{\partial n}{\partial\mu}\right|_{\mu \rightarrow 0^{+}} \simeq \sqrt{\frac{b}{2g}}\frac{1}{\mu^{1/2}}.\label{05}
\end{equation}
These equations ensure the divergence of the isothermal susceptibility at criticality. We point out that $\kappa_T \to \infty$ as $\mu \to 0^{\pm}$ (which is equivalent to $T \to T_c^{\pm}$) with the same exponent from both sides, as it should be \cite{Fisher-rpp30, Amit-book1984, Ma-book2000}.\\

While not a strict thermodynamic requirement, but to make the model as minimal as possible, we can require the derivative of $n$ to be continuous at the transition in order to fix the coefficient $b(T)$. That is, by equating the derivatives in Eqs. (\ref{04}) and (\ref{05}), one finds that
\begin{equation}
b(T) = \frac{4 \pi}{\lambda_T^3} \frac{a_s}{\lambda_T} .\label{bT}
\end{equation}
This identification completely determines the equation of state here proposed, in terms of thermodynamic quantities, such as $T$ and $\mu$, Planck constant $\hbar$ and atomic properties such as mass $m$ and atomic interactions $a_s$. \\

\begin{figure}[t!]
\begin{center}
\includegraphics[width=0.5\columnwidth]{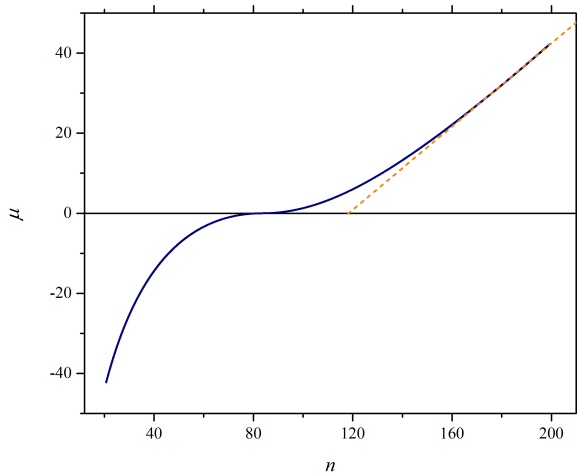}
\caption{(Color online) A typical isotherm with $T=9.5 \times 10^{-4} $ of the equation of state $\mu = \mu(n,T)$, as given by Eqs. (\ref{muneg}) and (\ref{mupos}). The second order phase transition occurs at $\mu = 0$. For $\mu \le 0$, the gas is represented by an ideal Bose gas, while for $\mu \ge 0$ it reaches asymptotically a Thomas-Fermi gas, indicated by the (orange) dotted line. Note that the curve is continuously flat at the transition, giving rise to the divergence of the isothermal compressibility. Units, $\hbar = m = a_s = 1$}
\label{Scheme}
\end{center}
\end{figure}

\begin{figure}[t!]
\begin{center}
\includegraphics[width=0.5\columnwidth]{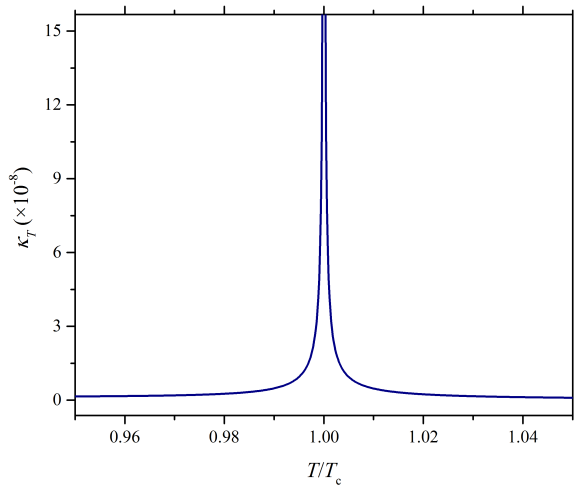}
\caption{Isothermal compressibility $\kappa_T$ as a function of $T$, at number density $n$ fixed, for a homogeneous gas. The divergent behaviour at $T_c$ yields a critical exponent $\gamma = 1$.  Units, $\hbar = m = a_s = 1$}
\label{kt-homo}
\end{center}
\end{figure}

The present model also guarantees that the critical exponents are those of the Spherical model, to which ideal BEC belongs \cite{Gunton-pr166}. To illustrate this fact, we calculate the isothermal compressibility, see Eq.(\ref{kappaT}), giving
\begin{equation}
\kappa_T (\mu,T) = \left\{
\begin{array}{ccc}
\frac{1}{g n^2} \frac{\mu + b g}{\sqrt{(\mu + b g)^2 - (b g)^2}} &{\rm if} & \mu > 0 \\
\frac{1}{n^2 kT} \frac{1}{\lambda_T^3} g_{1/2}(\mu/kT) &{\rm if} & \mu \le 0
\end{array} \right.
\end{equation}
where $n = n(\mu,T)$ is given by Eqs. (\ref{muneg}) and (\ref{mupos}). As shown by Eqs. (\ref{04}) and (\ref{05}), $\kappa_T$ diverges as $\mu \to 0^{\pm}$. This critical behavior is better seen if for a fixed value of the density $n$ one finds the behavior of the compressibility as a function of temperature $T$. We verify that it diverges as
\begin{equation}
\kappa_T \sim |T - T_c(n) |^{-1} \>\>\>{\rm for}\>\>\>n = {\rm constant} ,\label{kap-div}
\end{equation}
from above and below the transition, and with $T_c(n)$ given by Eq.(\ref{nc}). The corresponding critical exponent is $\gamma = 1$. Fig. \ref{kt-homo} shows the behaviour of $\kappa_T$ in the vicinity of $T_c$ for a fixed value of $n$. As we show below, the {\it global} isothermal compressibility does not diverge at BEC for the harmonically confined gas, yet its first derivatives do so.\\

\begin{figure}[t!]
\begin{center}
\includegraphics[width=0.5\columnwidth]{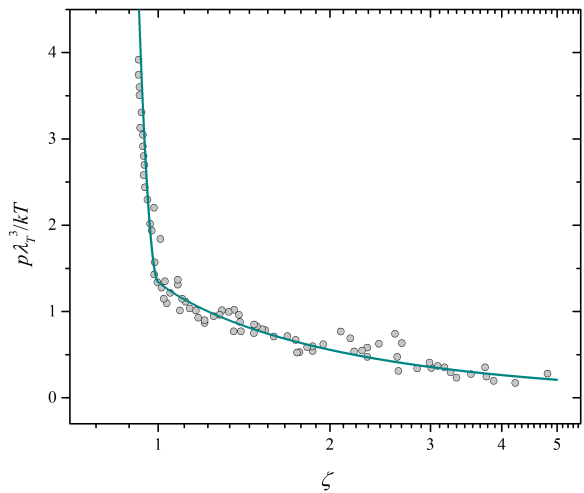}
\caption{(Color online) Plot of $p \lambda_{T}^3/kT$ as a function of $\zeta = e^{-\mu/kT}$ for an ultracold gas of $^{7}$Li atoms  at temperature $T = 1.6 \times 10^{-6}$ K, with scattering length $a_s = 8 \> a_0$, with $a_0$ Bohr radius. The experimental data was obtained from Nascimb\`ene et al. \cite{Nascimbene-njp12}. The continuous line is the present equation of state model, $p = p(\mu,T)$, Eqs. (\ref{pp1}) and (\ref{pp2}). There are no fitting parameters.}
\label{Pvszeta}
\end{center}
\end{figure}

Although the present equation of state is still a mean-field approximation, it may be of interest to compare it with actual experimental data. For this purpose, we recall the very accurate {\it in-situ} measurements of the atomic density profile $\rho(\vec r)$ by Nascimbene et al. \cite{Nascimbene-njp12}, on ultracold vapours of $^{7}$Li. In those experiments, using the local density approximation (LDA), described in the next section, the authors were able to extract the {\it homogeneous} pressure $p$ as a function of $\mu$ and $T$, as shown in Fig. \ref{Pvszeta}. To compare our equation of state $n(\mu,T)$, Eqs. (\ref{muneg}) and (\ref{mupos}), with the given experimental data, we need the equation $p = p(\mu,T)$. This can be found by integrating the Gibbs-Duhem relation $dp = n d\mu + s dT$ at constant $T$, yielding,
\begin{equation}
p(\mu,T) = \frac{kT}{\lambda_T^3} g_{5/2}(\mu/kT) \>\>\>{\rm if} \>\>\> \mu \le 0 \label{pp1}
\end{equation}
and 
\begin{eqnarray}
p(\mu,T) &=& \frac{kT}{\lambda_T^3} g_{5/2}(0) + n_c(T)\mu + \frac{b^2 g}{2} \left\{\left(1 + \frac{\mu}{bg}\right)\sqrt{\frac{\mu}{bg}\left(2 + \frac{\mu}{bg}\right)} - \right. \nonumber \\
&& \left. \ln \left[1 + \frac{\mu}{bg} + \sqrt{\frac{\mu}{bg}\left(2 + \frac{\mu}{bg}\right)} \right]\right\}
  \>\>\>{\rm if} \>\>\> \mu \ge 0 \label{pp2}
\end{eqnarray}
In Fig. \ref{Pvszeta} we also plot the above expression as a function of $\mu/kT$, using the measured temperature $T = 1.6 \times 10^{-6}$K and the scattering length $a_s = 8 a_0$, with $a_0$ the Bohr radius. The agreement is quite good given the fact that there are no further fitting parameters. However, in all fairness, other models \cite{Ketterle-proceeding1999,Nascimbene-njp12} may also provide acceptable agreement with experimental data. Those models usually assume that the density of the condensed phase is of the Thomas-Fermi type, namely, that it is given by $n =  \mu/g$. This assumption, while acceptable for data comparison purposes, yields a temperature independent isothermal compressibility, $\kappa_T = 1/n^2 g$, and, hence, a non-divergent compressibility as $T \to T_c$ from the condensed phase, at constant density $n$. 
Recalling that the compressibility is proportional to the density fluctuations, one finds that those models cannot describe the expected critical behavior of the superfluid BEC transition. As we described above in detail, the present model, given by Eqs. (\ref{muneg}) and (\ref{mupos}), has been heuristically built to have both, the TF behavior in the dense and low temperatures regions and, at the same time, yield the expected critical behavior at $T_c$. With these properties one can further study the critical behavior of the global thermodynamics of the confined Bose gas, as done in the following Sections.\\

In Section 4 we extend this discussion for the case of an inhomogeneous gas trapped in a harmonic potential, but first, in Section 3, we introduce the theoretical formalism that will be used for that goal, namely, the Global Thermodynamics.

\section{Global Thermodynamics of a confined fluid}

This section is devoted to a brief review of the global thermodynamics of a fluid confined by an inhomogeneous external potential $V_{ext}(\vec r)$ \cite{Romero-Rochin-prl94, Romero-Rochin-bjp35, Sandoval-Figueroa-pre78}. We call it ``global'' to make a distinction from the usual textbook descriptions of thermodynamics in which the usage of the mechanical variables volume $V$ and pressure $p$ is taken for granted. However, these variables are the appropriate ones only if the confining potential is a box of rigid walls. Once the system is confined by an external potential $V_{ext}(\vec r)$, the variables $V$ and $p$ (constant) are no longer thermodynamic variables. First, there are no walls to confine the system, and second, an external potential, such as a harmonic one, can extend itself (theoretically) to all space and so can the fluid too. Moreover, the presence of the potential is now felt throughout the fluid, and not only at the walls as in the homogenous case. This causes the usual volume $V$ to be replaced by an analogous variable that determines the spatial extent of the fluid. In addition, the presence of an inhomogeneous potential also causes the hydrostatic pressure to be no longer a constant throughout the fluid, as Pascal law indicates. This qualitative discussion, however, can be verified by appealing to the basic formulae of statistical physics, as we now review. \\

Consider a generic fluid of identical atoms interacting pairwise. The Hamiltonian is,
\begin{equation}
H = \sum_{i=1}^N \frac{\vec p^2}{2m} + \sum_{i < j} u(r_{ij}) + \sum_{j = 1}^N V_{ext}(\vec r_j) ,\label{H}
\end{equation}
where $u(r_{ij})$ is a two-body short-range isotropic potential. Thermodynamics is given by the free energy $F = - kT \ln Z$, for a given temperature, with the partition function given by,
\begin{equation}
Z = {\rm Tr} \>e^{-H/kT} .
\end{equation}
The thermodynamic limit should be taken and this gives rise to the identification of the mechanical thermodynamic variables appropriate to the confining external potential $V_{ext}(\vec r)$. For the sake of argument, let us consider an anisotropic harmonic potential $V_{ext}(\vec r) = m (\omega_x^2 x^2 + \omega_y y^2 + \omega_z^2 z^2)/2$. The partition function $Z$ is then a function of $N$, $T$, of the frequencies $\omega_x$, $\omega_y$, $\omega_z$, and of parameters of the interatomic potential $u(r_{ij})$. For simplicity, one can first analyze ideal gases, namely, neglecting the potential $u(r_{ij})$. An straightforward calculation of a classical ideal gas, after considering $N \gg 1$, yields,
\begin{equation}
F(N,T) = -N kT \left(\ln \frac{1}{N} \left(\frac{kT}{\hbar \bar \omega}\right)^3 + 1\right) .
\end{equation}
where $\bar \omega = (\omega_x \omega_y \omega_z)^{1/3}$ is the mean of the trap frequencies.
In the same way, the calculation of the grand potential of an ideal Bose gas obtains,
\begin{equation}
\Omega(\mu,T) = -kT \left(\frac{kT}{\hbar \bar \omega}\right)^3 g_4(\mu/kT)
\end{equation}
where $g_4(\mu/kT)$ is the $n = 4$ Bose function, Eq.(\ref{g}). First of all, one observes that the variable volume $V$ is absent. This is because the gas is confined by a harmonic trap and not by a bottle or box of volume $V$. On the other hand, in the thermodynamic limit all extensive quantities $N$, $F$, $\Omega$, $S$, and so on, must diverge, with their ratios remaining constant, namely $\Omega/N$, $S/N$, etc, finite. Hence,  from the above expressions one must conclude that, in order to preserve extensitivity, $ \bar \omega^3 \to 0$, such that $N \bar \omega^3$, $\Omega  \bar \omega^3$, $F \bar \omega^3$, and so on remain constant. This suggests the definition of an extensive ``global volume'' as, \cite{Groot-procrslona203,Romero-Rochin-prl94, Romero-Rochin-bjp35, Sandoval-Figueroa-pre78}.
\begin{equation}
{\cal V} = \frac{1}{ \bar \omega^3} .
\end{equation}
One can further analyze an atomic interacting fluid and reach the same conclusion \cite{Sandoval-Figueroa-pre78}. Additionally, one can prove that if ${\cal V}$ is changed adiabatically, for a fixed number of atoms $N$, the system either cools down or heats up. That is,  ${\cal V}$ is a bona-fide thermodynamic variable. It then follows that the free energies depend on it as $F = F(N, {\cal V},T)$ and $\Omega = \Omega({\cal V},T,\mu)$, in complete analogy to their dependence on the volume $V$ of a box. The interpretation is that, depending on the particular confining potential, this manifests itself through an extensive, mechanical variable that can be generally called a ``volume". Certainly, if the temperature and chemical potential are fixed, an increase in ${\cal V}$ must be accompanied by an increase in $N$, namely, it must behave as a ``volume'', regardless of its units. Moreover, any ``volume'' must have its conjugate ``pressure'', defined as, 
\begin{equation}
{\cal P} = - \left(\frac{\partial F}{\partial {\cal V}}\right)_{N,T}
\end{equation}
such that the reversible mechanical work, in expanding or contracting the confining harmonic trap, is given by $\dbar W = - {\cal P} d{\cal V}$, which indeed it is. A useful, general, formula for ${\cal P}$ is given by,
\begin{equation}
{\cal P V} = \frac{1}{3} \int \rho(\vec r) \vec r \cdot \nabla V_{ext}(\vec r) d^3 r, \label{PG}
\end{equation}
where $\rho(\vec r)$ is the average particle density, which is inhomogeneous due to the external potential. This is actually the density profile measured in experiments with ultracold gases. The above formulae are valid whether the system is classical or quantum \cite{Sandoval-Figueroa-pre78}. The particle density is an intensive quantity and, thus, it can depend on, say, $(N/{\cal V}, T)$ or $(\mu,T)$, depending on which ensemble the average is taken. Certainly, the number of particles is given by,
\begin{equation}
N = \int \> \rho(\vec r) \> d^3 r .\label{NG}
\end{equation}

Although the variables ${\cal P}$ and ${\cal V}$ do not have units of pressure and volume, respectively, the important point is that these are the bona-fide mechanical variables that replace $p$ and $V$ of the homogeneous case. While the global volume can be seen to represent the available physical space,  for given $N$ and $T$, as discussed above, the global pressure is more difficult to interpret. Yet, a closer look at the last expression of Eq. (\ref{PG}) indicates that the product ${\cal PV}$ not only has the units of energy, the right hand side is nothing but the virial expression for the ``pressure" \cite{Rowlinson-book2013}. That is, if we replace the harmonic potential by one of a vessel of rigid walls, then ${\cal P}$ and ${\cal V}$ are replaced by $p$ and $V$ in Eq. (\ref{PG}) and this expression becomes the usual one for the virial expansion of the pressure \cite{Sandoval-Figueroa-pre78}. Nevertheless, because the variables ${\cal P}$ and ${\cal V}$ are still unfamiliar, it may be useful to grasp their order of magnitude in a typical experiment with $^{87}$Rb at $T = 100 \times 10^9$ K, with $N \simeq 10^5$ atoms in a trap of mean frequency $\bar \omega = 2 \pi (100)$ Hz. The generalized volume is ${\cal V} \simeq  4.0 \times 10^{-9}$ s$^3$ and a typical generalized pressure ${\cal P} \simeq   5.2 \times 10^{-5}$ J s$^{-3}$. The product ${\cal P V} \simeq  1.2 \times 10^{-25}$ J, which compared with $N k T \simeq  1.4 \times 10^{-25} $J is of the same order, but very large compared with the harmonic quantum of energy $\hbar \bar \omega \simeq  6.6 \times 10^{-32}$ J, as it should be. Regarding typical sizes, this trap  corresponds to a mean harmonic length $(\hbar /m\bar \omega)^{1/2} \simeq  1.1 \times 10^{-6}$ m and a Thomas-Fermi radius of a condensate $R_{TF} \simeq  3.8 \times 10^{-5}$ m. \\

For a harmonic potential, the expression (\ref{PG}) gives us a tool to calculate the global pressure from the knowledge, by theory or experiment, of the density profile $\rho(\vec r)$, 
\begin{equation}
{\cal P} = \frac{2}{3{\cal V}} \int \> \rho(\vec r) \> \left( \frac{1}{2} m\omega^2 r^2 \right) \> d^3r .
\end{equation}
In turn, this expression allows us to calculate essentially all thermodynamic properties, in particular those related to the critical transition here discussed. For instance, for given ${\cal V}$, $T$ and $N$, see Eq. (\ref{NG}), one can construct the equation of state of the confined fluid in the form ${\cal P} = {\cal P}(N/{\cal V},T)$. With this, one calculates the global isothermal compressibility,
\begin{eqnarray}
{\cal K}_T & = & - \frac{1}{\cal V} \left(\frac{\partial {\cal V}}{\partial {\cal P}}\right)_{N,T} \nonumber \\
& = & \frac{\cal V}{N^2} \left(\frac{\partial N}{\partial \mu}\right)_{{\cal V},T} \label{KT} .
\end{eqnarray}
where, the second form follows from a thermodynamic identity, which will be very useful below. This expression is not an analogy to the homogeneous one $\kappa_T$, Eq. (\ref{kappaT}), it is the true susceptibility of the system to a change in the external potential, namely, a change in the trapping frequency $\omega$, and it is further related to the density fluctuations in the same way as $\kappa_T$ is related to density fluctuations in the homogeneous case. It is also amenable to be measured \cite{Poveda-Cuevas-pra92}. One can also calculate heat capacities and any other thermodynamic function of the confined system. We defer those and their analysis around criticality to the next section. \\

Now we turn to local density approximation (LDA) as a theoretical tool to obtain the density profile $\rho(\vec r)$, which is the crucial quantity for the obtention of global thermodynamic variables. LDA provides the bridge between the homogeneous and non-uniform versions of the same system, subject to the corresponding external fields. It can be shown to be exact in the thermodynamic limit \cite{Marchioro-commmathphys27, Marchioro-commmathphys29, Sandoval-Figueroa-pre78}. Its recipe is quite simple: one first obtains the homogeneous equation of state of the particle density as a function of chemical potential and temperature, $n = n(\mu, T)$; then, LDA proceeds replacing the chemical potential by a ``local" one, $\mu \to \mu - V_{ext}(\vec r)$ and, as a result, $n$ becomes the density profile $\rho(\vec r;\mu,T)$ of the inhomogeneous system
\begin{equation}
\rho(\vec r;\mu,T) = n(\mu - V_{ext}(\vec r),T). 
\end{equation}
Care should be taken that one obtains the density profile but at constant $\mu$ and $T$. All the thermodynamic identities follow, just at those  thermodynamic variables given. It is important to mention that, although LDA provides the desired bridge between the homogeneous and the inhomogeneous cases, with the chemical potential playing a very important theoretical role, in practice one does not need the specific knowledge of the value of the chemical potential. That is, knowledge of the density profile $\rho(\vec r)$ and the temperature $T$ suffice. With these, one finds the number of particles $N$ and the pressure ${\cal P}$, via Eq. (\ref{PG}), and the equation of state of state ${\cal P} = {\cal P}(N/{\cal V},T)$ follows.

\section{Critical thermodynamics of a trapped gas}

In this section we report the global critical thermodynamics of a harmonically confined Bose gas, using LDA with homogeneous equation of state introduced in Section II. Hence, restricting ourselves to an isotropic harmonic trap of frequency $\omega$, $V_{ext}(\vec r) = m \omega^2 r^2/2$, the density profile follows within LDA, 
\begin{equation}
\rho(r;\mu,T) =
\frac{1}{g}\sqrt{\left(\mu-\frac{1}{2}m\omega^{2}r^{2}+bg\right)^{2}-\left(bg\right)^{2}} + n_{c}(T)  \label{belowRTF}
\end{equation}
 if $\mu-\frac{1}{2}m\omega^{2}r^{2}\geq 0$, and 
\begin{equation} 
\rho(r;\mu,T) = \frac{1}{\lambda^{3}}g_{\frac{3}{2}}\left({\frac{\mu-\frac{1}{2}m\omega^2 r^{2}}{kT}}\right) \label{aboveRTF}
\end{equation}
if $\mu-\frac{1}{2}m\omega^{2}r^{2} \le 0$. \\

\begin{figure}[t!]
\begin{center}
\includegraphics[width=0.5\columnwidth]{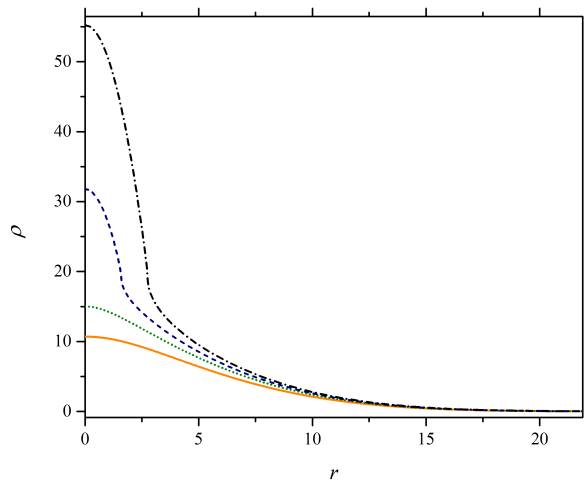}
\caption{(Color online) Density profile $\rho (r)$ as a function of $r$, for different values of the chemical potential. Solid line $\mu=-0.71$  and dotted line $\mu=-0.35$ are below BEC transition, dashed line $\mu=0.71$ and dot-dashed line $\mu=2.08$ above the transition. The temperature is $kT = 36.42$. Units, $\hbar = m = \omega = 1$.}
\label{dens-T}
\end{center}
\end{figure}

The consideration of an isotropic trap does not affect the final thermodynamic results, it simply makes the calculations easier. The above expressions indicate that the onset of BEC for an inhomogeneous trapped fluid occurs also for zero chemical potential, $\mu = 0$. That is, if $\mu < 0$, the profile is given by the ideal contribution only, Eq.(\ref{aboveRTF}), and the state is a normal thermal cloud. As the chemical potential changes sign, $\mu > 0$, the profile as a function of $r$ is then given by both equations Eqs. (\ref{belowRTF}) and (\ref{aboveRTF}), the former representing the condensate superfluid ``peak" and the latter the surrounding thermal cloud. The matching point of both solutions is found at $r = R_{TF}$ where 
\begin{equation}\label{RTF}
R_{TF}=\sqrt{\frac{2\mu}{m\omega^{2}}}
\end{equation}
has the usual form of the Thomas-Fermi radius. The interesting difference here with respect to its common identification is that it indeed indicates the spatial location of the condensate, but for arbitrary temperatures and not just for zero temperature \cite{Dalfovo-rmp99,Pethick-book2002}. The BEC transition can also be seen from the singular behavior of the global thermodynamic variables, as we show below. We recall here that if we fix the number of particles $N$ in the trap, using Eq. (\ref{NG}), then the critical temperature $T_c$ can be found when $\mu = 0$. This is illustrated in Fig. \ref{dens-T} where we plot several density profiles for a fixed value of $N$ and we observe the onset of BEC as a function of chemical potential (equivalent to changing temperature $T$). The appearance of the condensate occurs at $T_c$ defined by $\mu = 0$, which implies that $R_{TF} \ne 0$ below that temperature.

\newpage

As discussed in the previous section, using the density profile, one can find the equation of state ${\cal P} = {\cal P}(N/{\cal V},T)$ and, hence the global isothermal compressibility ${\cal K}_T$, see Eq. (\ref{KT}). However, with the assumed spherical symmetry of the density profile, most of the global thermodynamic expressions can be quite simplified. We just give here the explicit form of the isothermal compressibility ${\cal K}_T$, see Eq. (\ref{KT}),
\begin{eqnarray}
{\cal K}_T &=& \frac{\cal V}{N^2} \int \left(\frac{\partial \rho(\mu - m \omega^2 r^2/2, T)}{\partial \mu }\right)_T d^3 r \nonumber \\
& = & \frac{{\cal V}^2}{N^2} \frac{4 \pi \omega}{m} \int_{0}^{\infty}  \rho(r; \mu , T) dr . \label{eq:20}
\end{eqnarray}
This simple expression for the global compressibility is very useful since it avoids taking explicit derivatives as indicated above. Further expressions for the relevant thermodynamic quantities can be found in the Appendix. \\

Fig. \ref{kvsT} shows a typical curve of the global isothermal compressibility ${\cal K}_T$ as a function of $T$. We see immediately the contrast with respect to the compressibility of the uniform counterpart at the critical temperature, see Fig. \ref{kt-homo}. That is, the latter shows its characteristic divergence, while the global one presents a change of curvature at the transition  only. This change of curvature, however, hides the singular behavior of the free energy which emerges through the derivative of the global compressibility. \\

\begin{figure}[t!]
\begin{center}
\includegraphics[width=0.5\columnwidth]{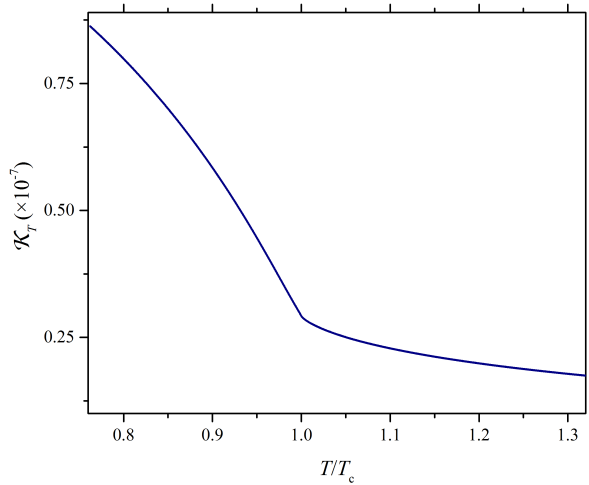}
\caption{A typical global isothermal compressibility ${\cal K}_T$, as a function of temperature $T/T_c$, for $N=5\times 10^{5}$ atoms in the trap. At the critical temperature $T_c$, the compressibility changes its curvature. As discussed in the text, the derivative of ${\cal K}_T$  becomes {\it logarithmically} divergent at $T_c$. Units, $\hbar = m = \omega = 1$.}
\label{kvsT}
\end{center}
\end{figure}

Although we are more explicit below, the behavior of ${\cal K}_T$, very near $T_c$ can be shown to be of the form
\begin{equation}
{\cal K}_T \approx {\cal K}_{T_c} + K_{\pm} \vert T-T_c \vert \left(\ln \vert  T-T_c \vert -1\right) + \dots
\end{equation}
where $K_{\pm}$ is a coefficient corresponding to $T \to T_c^{\pm}$, with $K_+ > 0$ and $K_- < 0$. The above expression indicates that the derivative of ${\cal K}_T$ with respect to $T$ diverges logarithmically at $T_c$, as shown by Fig. \ref{DerkT}. As can be seen in the Appendix, since we have access to analytic expressions for essentially all thermodynamic properties, as $\mu \to 0^+$, the derivative of ${\cal K}_T$ may be written as 
\begin{equation}\label{eq:2}
\left(\frac{\partial\mathcal{K}_{T}}{\partial T}\right)_{N,\mathcal{V}}\approx -k\frac{3kT}{\hbar^{3}\left(\frac{N}{\mathcal{V}}\right)^{2}}\frac{\tilde{g}_{3}(\alpha)}{\tilde{g}_{2}(\alpha)}\tilde{g}_{1}(\alpha)
\end{equation}
with $\alpha = \mu/kT$, and in which the only divergent term is $\tilde{g}_{1}$. On the other hand, if we consider $\mu \to 0^{-}$, the derivative has the form
\begin{equation}\label{eq:4}
\left(\frac{\partial\mathcal{K}_{T}}{\partial T}\right)_{N,\mathcal{V}}\approx -k\frac{3kT}{\hbar^{3}\left(\frac{N}{\mathcal{V}}\right)^{2}}\frac{{g}_{3}(\alpha)}{{g}_{2}(\alpha)}{g}_{1}(\alpha) .
\end{equation}
By expanding the above expressions near zero, and since $\tilde g_\nu \to g_\nu$, we obtain very near the transition,
\begin{equation}\label{eq:5}
\left(\frac{\partial\mathcal{K}}{\partial T}\right)_{N,\mathcal{V}} 
\approx -k\frac{3kT} {\hbar^{3}\left(\frac{N}{\mathcal{V}}\right)^{2}} \frac{\zeta(3)}{\zeta(2)}\ln(\vert T-T_{c}\vert)
\end{equation}
which shows the logarithmic divergence near $T_c$. This particular logarithmic divergence, while indicates the non-analytic nature of the transition, is a direct inheritance of the divergence of the isothermal compressibility $\kappa_T$ of the uniform system which, in this case, has the critical exponent $\gamma = 1$. In a real experimental gas the actual divergence of the derivative of the global isothermal compressibility should be in accord with the 3D XY model, which we do not expect to show a logarithmic divergence. We recall that a zero critical exponent is associated to a logarithmic divergence \cite{Fisher-rpp30, Amit-book1984, Ma-book2000}. 
We notice that the form of the peak in the derivative of the global compressibility near the critical point has remarkable similarities to the behavior of the isothermal compressibility for liquid helium as observed across the $\lambda$-transition, see Refs. \cite{Grilly-pr149, Boghosian-pr152, Elwell-pr164}.\\

\begin{figure}[t!]
\begin{center}
\includegraphics[width=0.5\columnwidth]{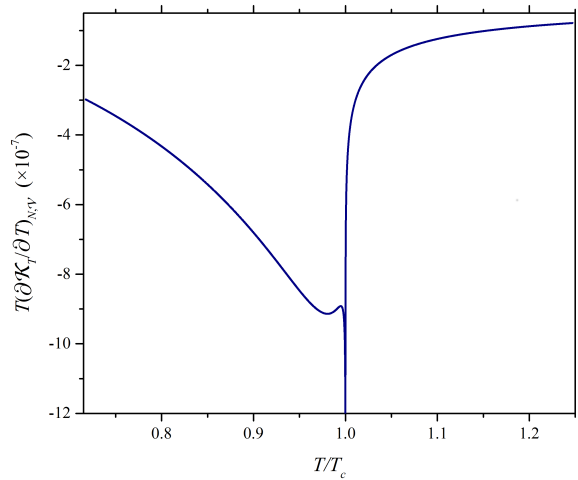}
\caption{Derivative of the isothermal compressibility, at constant $N$ and ${\cal V}$, $T \left(\frac{\partial {\cal K}_{T}}{\partial T}\right)_{N,{\cal V}}$as a function of temperature $T/T_c$, for $N=5\times 10^{5}$ atoms in the trap. As explained in the text, this derivative diverges logarithmically as $T \to T_c$. Units, $\hbar = m = \omega = 1$.}
\label{DerkT}
\end{center}
\end{figure}

\begin{figure}[t!]
\begin{center}
\includegraphics[width=0.5\columnwidth]{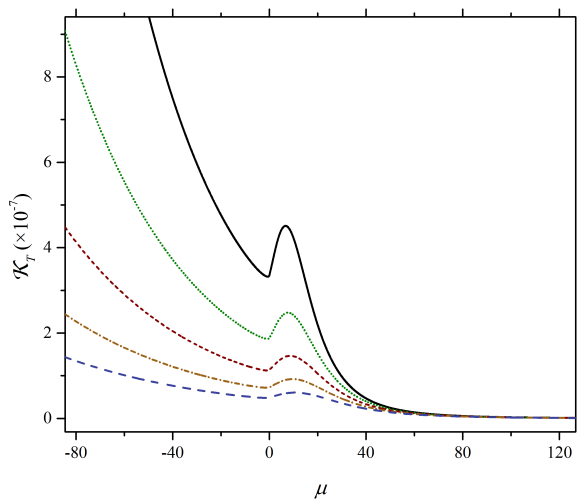}
\caption{(Color online) Global isothermal compressibility ${\cal K}_T$ as a function of chemical potential $\mu$, for several temperatures. Solid line $kT=42.0$, dotted line $kT=48.7$, dashed line  $kT=55.4$, dot-dashed line $kT=62.0$, bold-dashed line $kT=68.7$. Units, $\hbar = m = \omega = 1$. }
\label{kvsrho}
\end{center}
\end{figure}

Before analyzing other thermodynamic properties that also show critical behavior, such as the global specific heats and the coefficient of thermal expansion, we present Fig. \ref{kvsrho} showing several curves of the global isothermal compressibility as a function of global density $N/{\cal V}$ for different values of the global pressure ${\cal P}$. These plots are very similar to those recently reported by Bagnato et al. \cite{Poveda-Cuevas-pra92} obtained from measurements on a $^{87}\mathrm{Rb}$ BEC. Again, there appear peaks at the transition line but these are not divergent. \\

\begin{figure}[t!]
\begin{center}
\includegraphics[width=0.5\columnwidth]{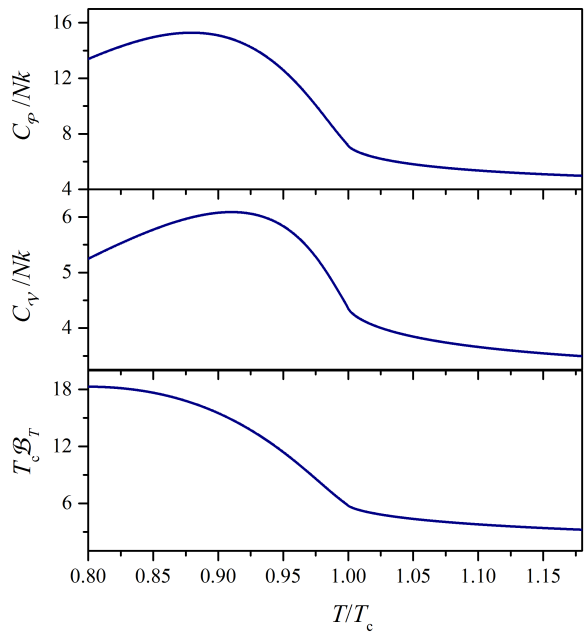}
\caption{Global specific heat at constant global pressure $C_{\cal P}/Nk$ (upper), global specific heat at constant global volume (middle), and global coefficient of thermal expansion $T_c {\cal B}_T$ (upper), as functions of  $T/T_c$. See text for discussion. The plotted variables are dimensionless.}
\label{three}
\end{center}
\end{figure}

We now turn our attention to three other quantities, relevant in the description of a phase transition. One is the heat capacity at constant global volume $C_{\cal V}$, other the heat capacity at constant global pressure $C_{\cal P}$ and the last one, the thermal global expansion coefficient ${\cal B}_T$. The formulae, respectively, are
\begin{equation}
C_{\cal V} = T\left(\frac{\partial S}{\partial T}\right)_{N,{\cal V}} \label{CV}
\end{equation}
\begin{equation}
C_{\cal P} = T\left(\frac{\partial S}{\partial T}\right)_{N,{\cal P}} \label{CP}
\end{equation}
\begin{equation}
{\cal B}_T = \frac{1}{\cal V} \left(\frac{\partial {\cal V}}{\partial T}\right)_{N,{\cal P}} \label{BT}
\end{equation}
The panel of Fig. \ref{three} shows the behavior of these quantities as a function of temperature $T$, in the vicinity of a critical temperature $T_c$. The first observation is that the three of them are finite at $T_c$, and the three of them also show a change of curvature at the transition in the same way as the isothermal compressibility does, see Fig. \ref{kvsT}. Therefore, their respective derivatives of $C_{\cal P}$ and ${\cal B}_T$ with respect to temperature also diverge logarithmically at the transition. This is expected based on the identity,
\begin{equation}
C_{\cal P} - C_{\cal V} = {\cal V} T \frac{{\cal B}_T^2}{{\cal K}_T}
\end{equation}
analogous to the one relating their homogenous counterparts \cite{Fisher-rpp30}. That is, the critical behavior of $C_{\cal P}$, ${\cal B}_T$ and ${\cal K}_T$ should be the same. The behavior of $C_{\cal V}$ is different from the previous ones, without divergences of neither the function itself nor its first derivatives. The fact that $C_{\cal V}$ has its own critical behavior, different from ${\cal K}_T$, $C_{\cal P}$ and ${\cal B}_T$, is because the latter are related to density fluctuations while $C_{\cal V}$ to energy fluctuations \cite{Landau-book2000}. We bring again the result of Ref. \cite{Donner-science315}, in which it was experimentally shown that the local correlations of a trapped BEC showed agreement with the XY-model. That is, we insist that such a model is still the underlying universality class of this phase transition, but modified by the presence of the trap. However, this point certainly needs further elucidation.\\

Following the previous paragraph, we point out that the maxima in $C_{\cal V}$ and $C_{\cal P}$ below $T_c$, see Fig. \ref{three}, should not be considered as precursors of a divergent behavior. Rather, we believe those maxima should be there in order to accommodate for the vanishing requirement of $C_{\cal V}$ and $C_{\cal P}$ as $T \to 0$. Nevertheless, these maxima, besides being experimentally testable, perhaps indicate a peculiar behavior that should be further addressed.
We recall that in Ref. \cite{Shiozaki-pra90} the measurement of $C_{\cal V}$ in a $^{87}$Rb vapor was reported with the corresponding plots very similar to those of Fig. \ref{three}.

\section{Final Remarks}

With the purpose of analyzing the critical behavior of a diluted non-uniform trapped ultracold gas across the BEC superfluid transition, we have introduced a model for the equation of state of a homogenous gas that shows a second order phase transition. The model assumes that in the normal gas states the behavior is fairly well represented by an ideal Bose gas, while in the BEC-superfluid states it asymptotically reaches the equation of state of a weakly interacting Thomas-Fermi gas. The matching is made at the transition where we impose continuity at the values of the chemical potential and of its first derivative  with respect to the particle density. With these requirements one finds the equation of state $n = n (\mu,T)$ of a homogenous Bose gas that presents a second order phase transition at BEC. This transition is in the spherical model universality class \cite{Fisher-rpp30} and, therefore, it cannot quantitatively describe the behavior of a real gas, which is expected to be in the 3D XY model class. Nevertheless, it serves to illustrate the main result of this article, namely, the fact that the critical behavior of an inhomogeneous trapped gas differs from its homogeneous counterpart.\\

Using the proper global thermodynamics of a trapped gas, and LDA approximation to obtain explicit expressions, we have shown that the critical behavior of a harmonically confined Bose gas across BEC does not show divergent thermodynamic susceptibilities but rather, typically continuous one. This continuity, however, does not indicate the lack of a singular behavior at a critical transition since, indeed, divergences appear now at the derivatives of the susceptibilities. This also does not seem to suggest that one is facing a different universality class from its homogenous partner. It appears that this is the manifestation of the critical behavior of the universality class of the homogenous system when this is confined by a particular trap. To be more precise, within the present model, the homogenous isothermal compressibility diverges with an exponent $\gamma = 1$, while the corresponding {\it global} isothermal compressibility does not diverge but its temperature derivative does so logarithmically. Due to the relationship among the homogenous thermodynamic variables and its corresponding global ones, as given by LDA, one can conclude that there is a one to one correspondence of the singular behavior. \\

We believe the present results, beyond the equation of state being useful as a fitting model for the density profile, points the way to perform ``high-resolution" measurements of density profiles and thermodynamic variables in order to elucidate the critical behavior of trapped gases. This also should go hand in hand with theoretical efforts to find an accurate equation of state that would lead to the correct critical exponents of the transition. \\

\section*{Acknowledgments}
We acknowledge support from grants CONACYT 232652, 260704, 271322, 254942 and 255573, PAPIIT-UNAM IA101716 and IN105217. I.R.A. thanks CONACYT for a graduate studies scholarship, and F.J.P.C. thanks SECITI 064/2015 and DGAPA-UNAM for postdoctoral fellowships.

\appendix

\section{Explicit expressions for global thermodynamic variables and their derivatives.}\label{AppendixkappaT}
 
As indicated in Eqs. (\ref{belowRTF}) and (\ref{aboveRTF}), the density profile for $\mu \le 0$ ($R_{TF} = 0$) is just the ideal Bose gas confined in a harmonic trap, while for $\mu > 0$ ($R_{TF} \ne 0$), one must use both forms of the profile. We obtain the following explicit expressions, for $\mu \le 0$,
\begin{equation}
N = {\cal V} \left(\frac{kT}{\hbar}\right)^3 \> g_3\left(\alpha\right)
\end{equation}
and for $\mu > 0$,
\begin{equation}\label{eq:12}
N\left(\mu,T,{\cal V}\right) = \frac{4 \pi }{3} \frac{R^3_{TF}}{\lambda^3} \left[ \frac{2 b \lambda^3}{5} Z_{1}\left( x \right) + \zeta\left(\frac{3}{2}\right) + \frac{3 \zeta \left(\frac{5}{2}\right)}{2 \alpha}+ \frac{3 \sqrt{\pi}}{4 \alpha^{3/2}} \tilde{g}_{3}\left(\alpha \right)\right],
\end{equation}
where
\begin{equation*}
Z_{1}\left(x\right) = \frac{\sqrt{2x+1}}{x} \left[\left(4x^{2}+2x+1\right)\mathrm{E}\left(\frac{1}{2x+1}\right)-x(4x+1)\mathrm{K}\left(\frac{1}{2x+1}\right)\right].
\end{equation*}
where $x=\frac{2\pi g^{2}}{\mu\lambda^{6}kT}$, $\mathrm{K}(x)$ and $\mathrm{E}(x)$ are the complete elliptical integrals of first and second kind, and where we have defined the function $\tilde{g_{s}}$ as,
\begin{equation}\label{eq:14}
\tilde{g}_{s}(\alpha)=\sum_{l=1}^{\infty}\frac{e^{l \alpha}}{l^{s}} \mathrm{Erfc}(\sqrt{l \alpha}),
\end{equation}

For the global pressure we obtain, for $\mu \le 0$,
\begin{equation}
{\cal P} = kT \left(\frac{kT}{\hbar}\right)^3 \> g_4\left( \alpha \right)
\end{equation}
and for $\mu > 0$, 
\begin{eqnarray}\label{eq:15}
\mathcal{P}\left(\mu,T\right)=& \frac{16 \sqrt{2} \pi }{3 m^{3/2}} \frac{\mu^{5/2}}{\lambda^3} \left[ \frac{2 b \lambda^3}{105 } Z_{2}\left(x\right) + \frac{\zeta\left( \frac{3}{2} \right)}{5}+\frac{\zeta \left( \frac{5}{2} \right)}{2 \alpha }+\frac{3 \zeta \left( \frac{7}{2} \right)}{4 \alpha^2}\right. \nonumber \\ &+ \left. \frac{3\sqrt{\pi}}{8 \alpha^{5/2} }\tilde{g}_{4}\left(\alpha \right) \right],
\end{eqnarray}
where
\begin{eqnarray}
Z_{2}\left(x\right) =& \frac{\sqrt{2x+1}}{x}\left[\left(32x^{3}+38x^{2}+9x+3\right)\mathrm{E}\left(\frac{1}{2x+1}\right) \right. \nonumber \\ &- \left. \left(32x^{3}+30x^{2}+3x\right)\mathrm{K}\left(\frac{1}{2x+1}\right)\right]. \nonumber
\end{eqnarray}

The global isothermal compressibility $\mathcal{K}_T$ is given as, for $\mu \le 0$,
\begin{equation}
 \mathcal{K}_T= \frac{1}{kT}\frac{{\cal V}^2}{N^2} \left(\frac{kT}{\hbar}\right)^3 \> g_2\left(\alpha \right)
\end{equation}
and for $\mu > 0$,
\begin{equation}
 \mathcal{K}_T=\frac{4\sqrt{2}\pi}{m^{3/2}}\frac{\mu^{1/2}}{\tilde{\rho}^{2} \lambda^3} \left[ \frac{2 b \lambda^3}{3} Z_{3}\left(x\right)  + \zeta\left(\frac{3}{2}\right)+\frac{\sqrt{\pi} }{2 \alpha^{1/2}}\tilde{g}_{2}\left(\alpha \right)\right], \label{eq:21}
\end{equation}
where
\begin{equation*}
Z_{3}\left(x\right) = \frac{\sqrt{2x+1}}{x} \left[\left(x+1\right) \mathrm{E}\left(\frac{1}{2x+1}\right)-x \mathrm{K}\left(\frac{1}{2x+1}\right)\right]
\end{equation*}

The derivative of the global isothermal compressibility requires the calculation of the following identity,
\begin{equation}\label{eq:22b}
\left(\frac{\partial{\cal K}_{T}}{\partial T}\right)_{N,{\cal V}}=\left(\frac{\partial {\cal K}_{T}}{\partial T}\right)_{\mu,{\cal V}}-\left(\frac{\partial{\cal K}_{T,{\cal V}}}{\partial\mu}\right)_{T}\frac{\left(\frac{\partial N }{\partial T}\right)_{\mu,{\cal V}}}{\left(\frac{\partial N}{\partial\mu}\right)_{T,{\cal V}}}.
\end{equation}
We calculate each derivative and we get the next expressions ($\mu\geq 0$):
\begin{eqnarray}\label{eq:23}
\left(\frac{\partial \mathcal{K}_T}{\partial T}\right)_{\mu} &= 
\frac{4 \sqrt{2} \pi }{m^{3/2}} \frac{\mu^{1/2} }{ \tilde{\rho}^{2} T \lambda^3} \left[ \frac{2 b \lambda^3}{3} Z_{4}\left(x\right)
+ \zeta\left(\frac{3}{2}\right)+\frac{\sqrt{\pi}}{\alpha^{1/2}} \tilde{g}_{2} \left(\alpha\right) \right. \nonumber
\\ &- \left. \frac{1}{2}\sqrt{\pi }\alpha^{1/2} \tilde{g}_{1}\left(\alpha\right)\right],
\end{eqnarray}
where
\begin{equation*}
Z_{4}\left(x\right)=\frac{1}{x \sqrt{2x+1}}\left[ \left(6 x^{2}+3 x \right)\mathrm{E}\left(\frac{1}{2x+1}\right)- 6x^{2} \mathrm{K}\left(\frac{1}{2x+1}\right)\right].\nonumber
\end{equation*}

\begin{equation}\label{eq:24}
\left(\frac{\partial \mathcal{K}_T}{\partial\mu}\right)_{T}=\frac{4 \sqrt{2} \pi}{m^{3/2} } \frac{1}{\tilde{\rho}^{2} \mu^{1/2} \lambda^{3}}
\left[ \frac{b  \lambda^3}{3 }Z_{5}\left(x \right)+ \zeta\left(\frac{3}{2}\right) + \frac{1}{2}\sqrt{\pi }\alpha^{1/2} \tilde{g}_{1}\left(\alpha\right) \right]
\end{equation}
with
\begin{equation*}
Z_{5}\left(x\right)=\frac{1}{x \sqrt{2x+1}}\left[ \left(x^2 +8 x +3\right)\mathrm{E}\left(\frac{1}{2x+1}\right)-\left(x^2 + 3x \right) \mathrm{K}\left(\frac{1}{2x+1}\right)\right]. \nonumber
\end{equation*}

\begin{eqnarray}\label{eq:25}
\left(\frac{\partial N }{\partial T}\right)_{\mu,\mathcal{V}}&= \frac{8 \sqrt{2} \pi }{m^{3/2}} 
\frac{\mathcal{V}\mu^{3/2}}{T \lambda^3} \left[ \frac{2 b \lambda^3}{3} Z_{6}\left(x\right) + \frac{\zeta\left(\frac{3}{2}\right)}{2} + \frac{3 \zeta \left(\frac{5}{2}\right)}{2\alpha} \right. \nonumber \\ &- \left. \frac{3\sqrt{\pi}}{4 \alpha^{1/2}} \tilde{g}_{2} \left(\alpha\right)   + \frac{3 \sqrt{\pi}}{4 \alpha^{3/2}} \tilde{g}_{3} \left(\alpha\right) \right]
\end{eqnarray}
where
\begin{eqnarray}
Z_{6}\left(x\right) &= \frac{1}{x \sqrt{2x+1}}\left[ \left(8x^{3} + 6 x^{2}+ x\right) \mathrm{E}\left(\frac{1}{2x+1}\right)\right. \nonumber \\ &- \left. \left(8x^{3}+4 x^{2} \right) \mathrm{K}\left(\frac{1}{2x+1}\right)\right] \nonumber
\end{eqnarray}

\begin{equation}\label{eq:26}
\left(\frac{\partial N}{\partial\mu}\right)_{T,\mathcal{V}}= \frac{8 \sqrt{2} \pi }{m^{3/2}} \frac{\mathcal{V} \mu^{1/2}}{ \lambda^3} \left[ \frac{b \lambda^3}{3} Z_{7}\left(x\right) +  \frac{\zeta\left(\frac{3}{2}\right)}{2} + \frac{\sqrt{\pi}}{4 \alpha^{1/2}} \tilde{g}_{2} \left(\alpha\right) \right]
\end{equation}
with
\begin{eqnarray}
Z_{7}\left(x\right) &= \frac{1}{x \sqrt{2x+1}}\left[\left(2x^{2}+3x+1\right)\mathrm{E}\left(\frac{1}{2x+1}\right)\right. \nonumber \\ &- \left. \left(2x^{2}+ x\right)\mathrm{K}\left(\frac{1}{2x+1}\right)\right]. \nonumber
\end{eqnarray}
If $\mu\leq 0$
\begin{equation}
\left(\frac{\partial \mathcal{K}_{T}}{\partial T}\right)_{N,\mathcal{V}}=k\frac{kT}{\hbar^{3}\left(\frac{N}{\mathcal{V}}\right)^{2}}\frac{1}{g_{2}(\alpha)}\left[2g_{2}(\alpha)g_{2}(\alpha)-3g_{1}(\alpha)g_{3}(\alpha)\right]
\end{equation}

For the global thermal expansion coefficient we obtain, for $\mu\leq0$,
\begin{equation}
\mathcal{\beta}_{T}=\frac{1}{T}\left[4\frac{g_{2}\left(\alpha\right)g_{4}\left(\alpha\right)}{g_{3}\left(\alpha\right)g_{3}\left(\alpha\right)}-3\right]
\end{equation}
and for $\mu > 0$ we need the next derivatives
\begin{eqnarray}
\left(\frac{\partial \mathcal{P}}{\partial T} \right)_{\mu} &= \frac{16 \sqrt{2} \pi }{3 m^{3/2}} \frac{\mu^{5/2}}{T\lambda^3}
\left[ \frac{2 b \lambda^3}{105} Z_{8}\left(x\right) + \frac{3 \zeta\left(\frac{3}{2}\right)}{10} +\frac{5 \zeta\left(\frac{5}{2}\right)}{4 \alpha} + \frac{3 \zeta\left(\frac{7}{2}\right)}{\alpha^2} \right. \nonumber \\ &- \left. \frac{3 \sqrt{\pi}}{8 \alpha^{3/2}} \tilde{g}_{3} \left(\alpha\right)+\frac{3 \sqrt{\pi}}{2 \alpha^{5/2}} \tilde{g}_{4} \left(\alpha\right) \right]
\end{eqnarray}
where
\begin{eqnarray}
Z_{8}\left(x\right) &= \frac{1}{x \sqrt{2x+1}}\left[\left(448 x^4 +588 x^3 + 224 x^2 + 21 x \right) \mathrm{E}\left(\frac{1}{2x+1}\right) \right. \nonumber \\ &- \left. \left( 448 x^4 + 476 x^3 + 126 x^2\right)  \mathrm{K}\left(\frac{1}{2x+1}\right)\right] \nonumber
\end{eqnarray}
\begin{equation}
\left(\frac{\partial \mathcal{P}}{\partial \mu} \right)_{T} = \frac{16 \sqrt{2} \pi }{3 m^{3/2}}\frac{\mu^{3/2}}{\lambda^3}
\left[ \frac{2 b \lambda^3}{105} Z_{9}\left(x\right) + \frac{\zeta\left(\frac{3}{2}\right)}{2} +\frac{3 \zeta\left(\frac{5}{2}\right)}{4 \alpha} + \frac{3 \sqrt{\pi}}{8 \alpha^{3/2}} \tilde{g}_{3} \left(\alpha\right) \right]
\end{equation}
where
\begin{eqnarray}
Z_{9}\left(x\right) &= \frac{1}{x \sqrt{2x+1}}\left[\left(84 x^3 + 84 x^2 + 42 x + \frac{21}{2} \right)\mathrm{E}\left(\frac{1}{2x+1}\right) \right. \nonumber \\ &- \left. \left(84 x^3 + 63 x^2 + \frac{21}{2} x\right) \mathrm{K}\left(\frac{1}{2x+1}\right)\right]. \nonumber
\end{eqnarray}
The global heat capacity $\mathcal{C_{\mathcal{V}}}$ is given as:
\begin{equation}
\mathcal{C}_{\mathcal{V}}= T \left( \frac{\partial S}{\partial T} \right)_{\mu,\mathcal{V}} = T\left(\frac{\partial^{2}\mathcal{P}}{\partial T^{2}}\right)_{\mu}-T\left(\frac{\partial^{2}\mathcal{P}}{\partial\mu\partial T}\right)_{T}\frac{\left(\frac{\partial N}{\partial T}\right)_{\mu}}{\left(\frac{\partial N}{\partial\mu}\right)_{T}}
\end{equation}
we need de next to derivatives for $\mu > 0$
\begin{eqnarray}
\left(\frac{\partial^{2} \mathcal{P}}{\partial T^{2}} \right)_{\mu} =& \frac{16 \sqrt{2} \pi }{3 m^{3/2}} \frac{\mu^{5/2}}{T^{2}\lambda^3}
\left[ \frac{2 b \lambda^3}{105} Z_{10}\left(x\right) + \frac{3 \zeta\left(\frac{3}{2}\right)}{20} +\frac{3 \zeta\left(\frac{5}{2}\right)}{2 \alpha} + \frac{9 \zeta\left(\frac{7}{2}\right)}{\alpha^2} \right.\nonumber \\
&+ \left. \frac{3 \sqrt{\pi}}{8 \alpha^{1/2}} \tilde{g}_{2} \left(\alpha\right) - \frac{9 \sqrt{\pi}}{4 \alpha^{3/2}} \tilde{g}_{3} \left(\alpha\right)+\frac{9 \sqrt{\pi}}{2 \alpha^{5/2}} \tilde{g}_{4} \left(\alpha\right) \right]
\end{eqnarray}

where:
\begin{eqnarray}
Z_{10}\left(x\right) =& \frac{1}{x (2x+1)^{3/2}}\\
&\times \left[\left(3584 x^{5}+5260 x^4 +2660 x^3 + 504 x^2 + 21 x \right) \mathrm{E}\left(\frac{1}{2x+1}\right)\right. \nonumber \\
&- \left.  \left(3584 x^{5}+ 4816 x^4 + 5544 x^3 + 336 x^2\right)  \mathrm{K}\left(\frac{1}{2x+1}\right)\right] \nonumber
\end{eqnarray}

\begin{eqnarray}
\left(\frac{\partial^{2} \mathcal{P}}{\partial \mu\partial T} \right)_{T} &= \frac{16 \sqrt{2} \pi }{3 m^{3/2}} \frac{\mu^{3/2}}{T\lambda^3}
\left[ \frac{ b \lambda^3}{105} Z_{11}\left(x\right) + \frac{3 \zeta\left(\frac{3}{2}\right)}{4} +\frac{9 \zeta\left(\frac{5}{2}\right)}{4 \alpha} \right. \nonumber \\  &- \left. \frac{3 \sqrt{\pi}}{8 \alpha^{1/2}} \tilde{g}_{2} \left(\alpha\right) + \frac{9 \sqrt{\pi}}{8 \alpha^{3/2}} \tilde{g}_{3} \left(\alpha\right) \right]
\end{eqnarray}
where
\begin{eqnarray}
Z_{11}\left(x\right) &= \frac{1}{x (2x+1)^{3/2}}\left[\left(1680 x^4 +2100 x^3 + 840 x^2 + 189 x \right) \mathrm{E}\left(\frac{1}{2x+1}\right)\right. \nonumber \\ &- \left. \left(1680 x^4 + 1680 x^3 + 420 x^2\right)  \mathrm{K}\left(\frac{1}{2x+1}\right)\right] \nonumber
\end{eqnarray}

and for $\mu\leq 0$ we obtain
\begin{equation}
\mathcal{C}_{\mathcal{V}}=\frac{3}{\hbar^{3}}\frac{\left(kT\right)^{4}}{Tg_{2}\left(\alpha\right)}\left[4g_{2}\left(\alpha\right)g_{4}\left(\alpha\right)-3g_{3}\left(\alpha\right)g_{3}\left(\alpha\right)\right]
\end{equation}

\section*{References}
\bibliographystyle{iopart-num}
\bibliography{bibhypmodel}

\end{document}